\documentstyle[prb,aps]{revtex}
\begin{document}
\draft
\tighten
\twocolumn[\hsize\textwidth\columnwidth\hsize\csname @twocolumnfalse\endcsname
\title{Quantum Interference in Superconducting Wire Networks 
and Josephson Junction Arrays: Analytical Approach based on 
Multiple-Loop Aharonov-Bohm Feynman Path-Integrals}
\author{Yeong-Lieh Lin$^{1}$ and Franco Nori$^{2,3,*}$}
\address
{
$^{1}$ Department of Physics, West Virginia University, Morgantown, West Virginia 26506-6315 \\
$^{2}$ 
Frontier Research System, The Institute of Physical 
and Chemical Research (RIKEN), Saitama 351-0198, Japan \\
$^{3}$ 
Center for Theoretical Physics, Physics Department, Center
for the Study of Complex Systems, The University of Michigan, Ann Arbor,
MI 48109-1120, USA ** 
}
\date{\today}
\maketitle
\begin{abstract}
We investigate analytically and numerically the mean-field
superconducting-normal phase boundaries of two-dimensional
superconducting wire networks and Josephson junction arrays
immersed in a transverse magnetic field.
The geometries we consider include square, honeycomb,
triangular, and {\it kagom\'e\/} lattices.  
Our approach is based on an analytical study of
multiple-loop Aharonov-Bohm effects: the quantum interference between
different electron closed paths where each one of them encloses a net
magnetic flux. Specifically, we compute exactly the sums of magnetic
phase factors, i.e., the lattice path integrals, on all closed lattice 
paths of different lengths. A very large number, e.g., up to $10^{81}$ 
for the square lattice, exact lattice path integrals are obtained.
Analytic results of these lattice path integrals then enable us to obtain 
the resistive transition temperature as a continuous function of the field.  
In particular, we can analyze measurable effects on the superconducting 
transition temperature, $T_c(B)$, as a function of the magnetic filed $B$, 
originating 
from electron trajectories over loops of various lengths.
In addition to systematically deriving previously observed features, 
and understanding the physical origin of the dips in $T_c(B)$ 
as a result of multiple-loop quantum interference effects, 
we also find novel results. In particular, we explicitly derive 
the self-similarity in the phase diagram of square networks. 
Our approach allows us to analyze the complex structure present in the 
phase boundaries from the viewpoint of quantum interference effects 
due to the electron motion on the underlying lattices.
The physical origin of the structures in the phase diagrams is derived 
in terms of the size of regions of the lattice explored by the electrons.  
Namely, the larger the region of the sample the electrons can explore
(and thus the larger the number of paths the electron can take),
the finer and sharper structure appears in the phase boundary.
Our results for {\it kagom\'e\/} and honeycomb lattices compare very well 
with recent experimental measurements by Ciao, Hues, Chaikin, Higgins, 
Bhattacharya, and Spencer (Phys. Rev. B, companion paper). 
\end{abstract}
\vskip2pc]
\narrowtext

\section{Introduction}

When immersed in an externally applied magnetic field,
superconducting networks\cite{1} made of thin wires, 
proximity-effect junctions, 
and tunnel junctions 
exhibit complex and interesting forms of phase diagrams.
These superconducting networks have been studied
in various kinds of geometries,
including simple\cite{1} and complex\cite{new5,5} periodic lattices,
regular fractals,\cite{6} bond-percolation networks,\cite{7}
disordered arrays\cite{8}
and quasiperiodic lattices.\cite{9,10,12,13,14}
The rich structure present in the
resistive transition temperature as a function of the magnetic field,
namely, the superconducting-normal phase diagram,
has a rich structure that has been the subject 
of various experimental and theoretical  
investigations.\cite{12,15,16,17,lin-kagome,loops}

\subsection{Physics of the Phase Diagram}

The rich structure in the phase diagram is essentially a result of
quantum interference effect or frustration due to the magnetic field
and the built-in  multi-connectedness of the networks.
The magnetic fluxes through
the cells of various areas, measured in units of the superconducting
flux quantum $\Phi_0\equiv hc/2e$, are useful parameters to characterize the
interference effect.  At zero magnetic field, the quantum interference effect
is absent, and therefore the resistive transition temperature
should have a peak.  Also, due to gauge invariance, physical quantities 
should be periodic functions of the cell fluxes, with a period of $\Phi_0$.
These arguments qualitatively explain the apparent periodic or
quasi-periodic structures observed in phase diagrams of networks
of various geometries.

To gain a quantitative description of the phase diagrams, we employ the
mean-field theory which is very effective in serving such a purpose.
For wire networks, the mean-field expression is given by the
Landau-Ginsburg equation expressed in terms of
the order parameters at the nodes.\cite{15}  For a junction array, one has
a set of self-consistent equations\cite{16,17} for the thermally
averaged pair wavefunctions of the grains.  Such equations
are linearized near the transition point, and the highest
temperature at which a non-trivial solution first appears is identified
as the transition temperature.  Therefore, one is left to find the top
spectral edge of eigenvalue problems.  The equations for a junction array
can be mapped into a tight-binding Schr\"{o}dinger problem
for an electron hopping on a lattice immersed in a magnetic field. The
equations for a wire network are in general more difficult to solve,
because the eigenvalue appears in a non-linear way. 

Numerical results
have been obtained for phase diagrams of networks of various geometries.
All of them compare very well with the corresponding experimental
data; the locations of the peaks of various sizes are correctly 
predicted and the relative heights of the peaks are also reproduced 
with occasional small deviations. The success of mean-field 
theory \cite{17,lin-kagome} suggests that much of the frustration effect 
in a statistical problem can be accounted for in terms of quantum
interference effect of linear wave mechanics.

\subsection{Many-loops generalization of the standard Aharonov-Bohm effect}

In this paper, we systematically investigate the field-dependent
superconducting-normal phase for a variety of two-dimensional
superconducting networks. The basis of our approach is the analytic
study of electron quantum interference effects originating from sums
over magnetic phase factors on closed lattice paths.
The sums of these phase factors, called lattice path integrals,
are many-loop generalizations of the
standard one-loop Aharonov-Bohm-type argument, where the electron
wave function picks up a phase factor $e^{i\Phi}$ each time it goes
around a closed loop enclosing a net flux $\Phi$. 

We compute analytically the lattice path integrals up to very long 
lengths for various types of lattices. These lattice path integrals 
contain the quantum interference of enormous numbers of closed paths.
Through an iterative approach, these results then enable us to
obtain the corresponding phase boundaries\cite{17,lin-kagome} as continuous
functions of the strength of the applied field. This method provides 
a systematic approximation scheme, through finite truncations, for 
the spectral edges of eigenvalue problems from which our mean-field 
phase diagrams can be computed. Thus, we can gain considerable theoretical
insight into the physical origin of the structure in the phase diagrams.
This approach also enables us to analyze the structure of the phase
boundaries from the viewpoint of the geometric features of the networks.
We apply this approach to study the phase boundaries of square,
honeycomb, triangular, and kagome lattices. Our studies provide
complete and detailed analysis of the relationship between the phase
diagram structures and the corresponding network geometries.

\subsection{Organization of the paper}

This paper is organized as follows. In Sec.~II, we describe the 
general formulation of our approach to the determination of phase
diagrams for a variety of periodic superconducting networks. To 
illustrate our calculational scheme, we first compute the 
Little-Parks oscillatory phase boundary of a single superconducting
loop in Sec.~III. In Sec.~IV, we apply this approach
to the superconducting square network.
We devote Sec.~V to the discussion of a very important and 
interesting feature observed in the phase boundary of
the square network, namely, the self-similarity. The superconducting 
honeycomb, triangular, and kagome networks are studied 
based on the same approach, respectively, in 
sections VI, VII, and VIII. In Sec.~IX, we discuss some general trends
in the application of this approach to these types of networks studied above.
Comparisons of the phase boundaries between a single superconducting loop
and the corresponding superconducting network are also made.
Furthermore, we present a brief discussion on the relationship between
our approach and other related methods.
In Sec.~X, we compare the phase boundaries of
honeycomb and kagome lattices. The last section summarizes our results.

\section{general formalism}

The physics of $T_{c}(B)$, the superconducting-normal
phase boundary as a function of the field $B$, is determined by
the electronic kinetic energy because the applied field induces 
a diamagnetic current in the superconductor.\cite{1} This current 
(proportional to the velocity)
determines the kinetic energy of the system.
In other words, the kinetic energy can be written in terms of
the temperature as
$$-\frac{{\hbar}^{2}}{2 m^{\ast}}
\mbox{\boldmath $\bigtriangledown$}^{2} \ \sim \
-\ \frac{{\hbar}^{2}}{2m^{\ast} {\xi(T)}^2} \ \sim \ T_c(B)-T_c(0),$$
where, for any superconductor, $m^{\ast}$ is twice the electron mass,
and
$$\xi(T) = \frac{\xi(0)}{\sqrt{1-T_c(B)/T_c(0)}},$$
is the temperature-dependent coherence length.
The problem of obtaining $T_{c}(B)$ is then
mapped to that of finding the spectral
edges of tight-binding electrons on the corresponding lattice.
Thus, assuming a unit hopping integral between adjacent sites, 
we consider the following Hamiltonian
\begin{equation}
H=\sum_{\langle ij \rangle}c_{i}^{\dag}c_{j}\exp(i A_{ij}),
\end{equation}
which describes the kinetic energy of electrons hopping on a discrete
lattice subject to a perpendicular magnetic field. Here $\langle ij \rangle$
refers to nearest-neighbor sites and the magnetic phase
$$ 
A_{ij}  =2\pi\int_{j}^{i}{\bf A}\cdot d{\bf l}
$$
is $2\pi$ times the line integral of the vector potential, ${\bf A}$,
along the bond from $j$ to $i$ in units of the $\Phi_0=hc/2e$.

\subsection{Sums over closed paths}

The lattice path integral, $\mu_{l}$, is defined as
\begin{equation}
\mu_{l}\equiv \sum_{\text{All closed lattice paths {\normalsize $\gamma$}
of length {\normalsize $l$}}} e^{i\Phi_{\gamma}}.
\end{equation}
By closed paths of length $l$ we mean the paths starting and ending at
the same site after traversing $l$ steps on the lattice
and $\Phi_{\gamma}$ is the sum over phases of the bonds on the path $\gamma$.
Let $|\Psi_i\rangle$ denote a localized electron state centered at site $i$.
It is not difficult to
notice that $\mu_{l}$ corresponds precisely to the quantum
mechanical expectation value $\langle \Psi_i |H^{l}| \Psi_i \rangle$,
which summarizes the contribution to the electron kinetic energy of
{\em all} closed paths of $l$-steps.

\noindent 
The physical meaning of the lattice path integral 
$$
\mu_{l} \ =\ \langle \Psi_i |H^{l}|\Psi_i \rangle
$$
thus becomes clear. The Hamiltonian $H$ is applied
$l$ times to the initial state $|\Psi_i\rangle$, resulting in the new
state 
$$ |\Psi_f\rangle  = H^{l}|\Psi_i\rangle $$ 
located at the end of the path traversing
$l$ lattice bonds. Because of the presence of a magnetic field, a
magnetic phase factor $e^{i A_{ij}}$ is acquired by an electron when
hopping from $j$ to the adjacent site $i$. 
The lattice path integral $\mu_{l}$ is nonzero only when 
the path ends at the starting site. In other words,
$\mu_{l}$ is the sum of the contributions from all {\em closed\/} paths
of $l$ steps starting and ending at the same site, each one weighted
by its corresponding phase factor $e^{i\Phi_\gamma}$ where
$$
\frac{\Phi_\gamma}{2\pi} \ = \ 
{\rm {\em net\/}\ flux\ enclosed\ by\ the\ closed\ path}\ \gamma.
$$

\subsection{Quantum Interference}

It is important to stress that $\Phi_{\gamma}$ depends crucially on the
traveling route of the path.\cite{17,lin-kagome}
 For instance, $\Phi_{\gamma}$ will be
positive (negative) by traversing a polygon loop counterclockwise
(clockwise). Therefore, {\it quantum interference\/} 
information contained in $\mu_{l}$ arises because 
the phase factors of different closed paths,
including those from all kinds of distinct loops and separate
contributions from the same loop, interfere with each other. Sometimes,
the phases corresponding to subloops of a main path cancel.

To analytically compute \cite{17,lin-kagome} 
the lattice path integrals $\mu_{l}$ is in general
a difficult task since $\mu_{l}$ involves an enormous number of
different paths (growing rapidly when $l$ increases), each one
determined by its corresponding net magnetic phase factor. We have developed
systematic and efficient methods to calculate the lattice path integrals
for a number of distinct lattices. The techniques involve successively
iterating the constructed recursion relations and exploiting the symmetries
of the underlying lattices. The technical details of the implementation 
will be presented elsewhere.
Below we will only list the first few calculated lattice path integrals
in relevant places. Results for the lattice path integrals of larger $l$
will not be presented due to their lengthy expressions, but will be 
used in some of our calculations.

In summary, the lattice path integrals summarize the electron quantum 
interference effects originating from sums over magnetic phase factors 
on closed lattice paths.  The sums of these phase factors, the lattice 
path integrals, are many-loop generalizations of the
standard one-loop Aharonov-Bohm-type argument, where the electron
wave function picks up a phase factor $e^{i\Phi}$ each time it goes
around a closed loop enclosing a net flux $\Phi$.

\subsection{Computation of the energy eigenvalues from lattice
path integrals}

We now outline the scheme for obtaining the eigenvalues from the calculated
lattice path integrals.
Let us apply the Hamiltonian to the starting state
$$ |\psi_1 \rangle \equiv |\Psi_{i} \rangle, $$
which is a localized state centered at an arbitrary site $i$
on the lattice, and perform the following expansions:
$$ H|\psi_1\rangle=a_{1}|\psi_1\rangle
+b_{2}|\psi_{2}\rangle$$
and for $n > 1$
$$H|\psi_n\rangle=b_{n}|\psi_{n-1}\rangle+a_{n}|\psi_n\rangle
+b_{n+1}|\psi_{n+1}\rangle.$$
The Hamiltonian matrix in the
basis $|\psi_n\rangle$ is obviously in a real tridiagonal form.
Each new state in this method expands outward by one more step from the site
where the starting state is located.
Note that the  $a_{n}$'s and $b_{n+1}$'s are gauge-invariant quantities.
Through these parameters we can construct the
truncated Hamiltonian matrices, $H^{(n)}$, which is the
$n$th-order approximation to $H$.
For instance,
\begin{eqnarray*}
H^{(2)} &=&\left[\begin{array}{cc}
a_{1} & b_{2}   \\
b_{2} & a_{2}
\end{array} \right], \\
H^{(3)} &=&\left[\begin{array}{ccc}
a_{1} & b_{2} & 0  \\
b_{2} & a_{2} & b_{3} \\
0 & b_{3} & a_{3}
\end{array} \right], \\
H^{(4)} &=&\left[\begin{array}{cccc}
a_{1} & b_{2} & 0 & 0  \\
b_{2} & a_{2} & b_{3} & 0  \\
0 & b_{3} & a_{3} & b_{4}  \\
0 & 0 & b_{4} & a_{4}
\end{array} \right],
\end{eqnarray*}
and so on. The quantity we desire, i.e., the top spectral edge
can then be obtained by solving the eigenvalues of $H^{(n)}$ and
will be designated by $T^{(n)}_{c}$, which is the $n$th-order
approximant to the phase boundary.
This scheme is useful because finite truncations give good approximations
to $T_{c}(B)$.

The coefficients $a_{n}$'s and $b_{n+1}$'s can be exactly
expressed in terms of the lattice path integrals in a systematic manner,
which will be presented below, respectively, for the bipartite and
non-bipartite lattices. In general, given the lattice path integrals up to
the order $\mu_{2L-1}$,
which contains information on the quantum interference effects due to
closed paths of $2L-1$ steps,
we can obtain the coefficients up to $a_{L}$ and $b_{L}$. Thus, the
$L$th-order truncation of the Hamiltonian matrix can be constructed,
and subsequently $T^{(L)}_{c}$ can be obtained.

\subsubsection{For bipartite lattices}

We first discuss the case for bipartite lattices where the lattice path
integrals of odd number steps are identically zero, i.e., 
$$\mu_{2l+1}=0.$$
It is evident that $$a_{n}=0$$ for any $n$.
To compute the $b_{n+1}$'s, we define an auxiliary
matrix $B$ with the first row elements given by
$$B_{1,l}\equiv \mu_{2l}.$$
The other rows are evaluated by using only one immediate predecessor
row. Namely, for $k \geq 2$ and $l \geq 1$
\begin{equation}
B_{k,l}=\frac{B_{k-1,l+1}}{B_{k-1,1}}
-\sum_{i=0}^{l-1}B_{k,i}B_{k-1,l-i},
\end{equation}
where $$B_{n,0}\equiv 1$$ for $n\geq 1$.
The $b_{n+1}$'s are obtained from the elements
of first columns of the matrix $B$ as
\begin{equation}
b_{n+1}=\sqrt{B_{n,1}}.
\end{equation}
Below we explicitly express the first few $b_{n+1}$'s in terms of
the lattice path integrals noting that $\mu_{2}$ is always equal to $z$,
the coordination number of the lattice:
\begin{eqnarray*}
b_{2} &=& \sqrt{\mu_{2}}=\sqrt{z}, \\
b_{3} &=& \sqrt{\frac{\mu_{4}}{z}-z}, \\
b_{4} &=& \sqrt{\frac{\mu_{6}-2\mu_{4}z+z^3}{\mu_{4}-z^2}
-\frac{\mu_{4}-z^2}{z}}.
\end{eqnarray*}
These expressions are applicable to any type of bipartite lattice.

It is worthwhile to point out that the number of elements on a specific row
is always less than that on the immediate predecessor row by one.
For instance, for a specific $k$, if the matrix elements run from
$B_{k,1}$ to $B_{k,l}$, the elements in the next row run from
$B_{k+1,1}$ to $B_{k+1,l-1}$. Therefore, given the lattice path
integrals up to $\mu_{2L}$, the matrix $B$ consist of $L$ rows.
The  $L$th (last) row has only one element $B_{L,1}$ from which 
we can deduce $b_{L+1}$. It is clear now that the highest-order 
approximation $T_{c}^{(L+1)}$ to the phase boundary can be 
obtained from $\mu_{2}, \mu_{4}, \ldots, \mu_{2L}$.

\subsubsection{For non-bipartite lattices}

Turning to the non-bipartite lattice case, we now define an auxiliary
matrix $N$ with the first row elements given by
$$N_{1,l}\equiv \mu_{l}.$$
The other rows are evaluated by using only one immediate predecessor
row. Namely, for $k \geq 2$ and $l \geq 1$
\begin{equation}
N_{k,l}=\frac{N_{k-1,l+2}-N_{k-1,1}N_{k-1,l+1}}{N_{k-1,2}
-N_{k-1,1}^{2}}-\sum_{i=0}^{l-1}N_{k,i}N_{k-1,l-i},
\end{equation}
where $N_{n,0}\equiv 1$ for $n\geq 1$. The $a_{n}$'s and $b_{n+1}$'s
are obtained from the elements
of the first and second columns as
\begin{equation}
a_{n}=N_{n,1}
\end{equation}
and
\begin{equation}
b_{n+1}=\sqrt{N_{n,2}-N_{n,1}^{2}}.
\end{equation}
Below we explicitly express the first few $a_{n}$'s and $b_{n+1}$'s
in terms of the lattice path integrals:
\begin{eqnarray*}
a_1 &=& 0, \\
a_2 &=& \frac{\mu{_3}}{z},   \\
a_3 &=& \frac{\mu_{5}z^{2}-2\mu_{4}\mu_{3}z+\mu_{3}^{3}}
{\mu_{4}z^{2}-\mu_{3}^{2}z-z^{4}},
\end{eqnarray*}
and
\begin{eqnarray*}
b_2 &=&\sqrt{z},  \\
b_3 &=&\sqrt{\frac{\mu_{4}}{z}-\frac{\mu_{3}^{2}}{z^2}-z}.
\end{eqnarray*}
The above expressions are valid for any type of non-bipartite lattice.

It is worth stressing that the number of elements on a specific row
is always less than that on the immediate predecessor row by two.
For instance, for a specific $k$ if the matrix elements runs from
$N_{k,1}$ to $N_{k,l}$, the elements in the next row runs from
$N_{k+1,1}$ to $N_{k+1,l-2}$. Therefore, given the lattice path
integrals up to $\mu_{2L+1}$, the matrix $N$ consist of $L+1$ rows.
The $L$th row has only three element $N_{L,1}$, $N_{L,2}$, and $N_{L,3}$,
where $b_{L+1}$ can be obtained from $N_{L,2}$, and $N_{L,3}$.
The $(L+1)$th (last) row has only one element $N_{L+1,1}$ from which we can
deduce $a_{L+1}$. It is clear now that the highest-order approximation
$T_{c}^{(L+1)}$ to the phase boundary can be obtained from
$\mu_{1}, \mu_{2}, \ldots,$ and $\mu_{2L+1}$.

\section{Simple illustration: a single superconducting loop}

Before we study the lattice cases, we apply the formalism described above
to three simple single-cell cases.
Namely, we calculate, respectively, the transition temperature of a single
superconducting loop in the shape of a square, a hexagon, and a triangle.
Exact solutions of the phase boundaries can be obtained for these simple
cases.  For all of these, $\Phi = \phi / 2\pi$ stands for the 
magnetic flux through these elementary cells, in units of $\Phi_{0}$.

The lattice path integrals, $\mu_{l}$, now correspond to the sums over all
closed paths of $l$ steps on a single cell. Closed-form results for the
lattice path integrals are derived. They are, respectively,
$$\mu_{2l}^{(s)}=C^{2l}_{l}+2\sum_{k=1}^{\left[l/2\right]}
C^{2l}_{l-2k}\cos(k\phi)$$
on a square,
$$\mu_{2l}^{(h)}=C^{2l}_{l}+2\sum_{k=1}^{\left[l/3\right]}
C^{2l}_{l-3k}\cos(k\phi)$$
on a hexagon, and
\begin{eqnarray*}
\mu_{2l}^{(t)} & = & C^{2l}_{l}+2\sum_{k=1}^{\left[l/3\right]}
C^{2l}_{l-3k}\cos(2k\phi), \\
\mu_{2l+1}^{(t)} & = & 2\sum_{k=0}^{\left[(l-1)/3\right]}C^{2l+1}_{l-3k-1}
\cos\left[(2k+1)\phi\right]
\end{eqnarray*}
on a triangle. Here
$$C^{m}_{n}=\frac{m!}{n!(m-n)!}$$
is the binomial coefficient, and the notation $[x]$ means the largest 
integer equal to or smaller than $x$. Through these results for the 
lattice path integrals, it is straightforward to compute the 
parameters $a_{n}$'s and $b_{n+1}$'s. 
In fact, for these small simple systems, the iterative process 
terminates very quickly.  In other words, the parameters 
$a_{n}$'s and $b_{n+1}$'s become identically zero after a few 
iterations. Hence, the corresponding exact tridiagonal Hamiltonian 
matrices can be readily constructed. 

\subsection*{Square loop}

Denoting the tridiagonal Hamiltonian matrix for the square
loop by $H_{s}$, we find that
$$H_{s}=\sqrt{2}\left[\begin{array}{cccc}
0 & 1 & 0 & 0  \\
1 & 0 & \left|\cos\left(\frac{\phi}{2}\right)\right| & 0  \\
0 & \left|\cos\left(\frac{\phi}{2}\right)\right| & 0 &
\left|\sin\left(\frac{\phi}{2}\right)\right|  \\
0 & 0 & \left|\sin\left(\frac{\phi}{2}\right)\right| & 0
\end{array} \right],$$
which is obtained by using only $\mu_{2}$, $\mu_{4}$, and $\mu_{6}$.
A closed-form expression for the top eigenvalue of $H_{s}$ 
can be easily obtained
$$T_{c}(\phi)=\sqrt{2+2\cos\left(\frac{\phi}{2}\right)}.$$

\subsection*{Hexagonal loop}

Similarly, denoting the tridiagonal Hamiltonian matrix for the hexagon
loop by $H_{h}$, we find that
$$H_{h}=\left[\begin{array}{cccccc}
0 & \sqrt{2} & 0 & 0 & 0 & 0  \\
\sqrt{2} & 0 & 1 & 0 & 0 & 0  \\
0 & 1 & 0  & \scriptstyle{\sqrt{1+\cos\left(\phi\right)}} & 0 & 0 \\
0 & 0  & \scriptstyle{\sqrt{1+\cos\left(\phi\right)}} & 0  &
\scriptstyle{\sqrt{1-\cos\left(\phi\right)}} & 0 \\
0 & 0 & 0 & \scriptstyle{\sqrt{1-\cos\left(\phi\right)}} & 0 & 1 \\
0 & 0 & 0 & 0 & 1 & 0
\end{array} \right],$$
which is obtained by using only $\mu_{2}$, $\mu_{4}$, $\mu_{6}$, $\mu_{8}$,
and $\mu_{10}$.
Let $j$ be an integer, the top eigenvalue of $H_{h}$ can be expressed as
follows:
$$T_{c}(\phi)=\left\{ \begin{array}{ll}
\sqrt{2+2\cos\left(\frac{\phi}{3}+\frac{2\pi}{3}\right)}  &
\mbox{for $\scriptstyle{
-\frac{3}{2}+3j \leq \frac{\phi}{2\pi} \leq -\frac{1}{2}+3j}$}  \\
2\cos\left(\frac{\phi}{6}\right) &
\mbox{for $\scriptstyle{
-\frac{1}{2}+6j \leq \frac{\phi}{2\pi} \leq \frac{1}{2}+6j}$}  \\
\sqrt{2+2\cos\left(\frac{\phi}{3}-\frac{2\pi}{3}\right)} &
\mbox{for $\scriptstyle{
\frac{1}{2}+3j \leq \frac{\phi}{2\pi} \leq \frac{3}{2}+3j}$} \\
-2\cos\left(\frac{\phi}{6}\right) &
\mbox{for $\scriptstyle{
\frac{5}{2}+6j \leq \frac{\phi}{2\pi} \leq \frac{7}{2}+6j}$}
\end{array} \right. . $$

\subsection*{Triangular loop}

Denoting the tridiagonal Hamiltonian matrix for the triangle
loop by $H_{t}$, we find that
$$H_{t}=\left[\begin{array}{ccc}
0 & \sqrt{2} & 0 \\
\sqrt{2} & \cos\left(\phi\right) & \left|\sin\left(\phi\right)\right| \\
0 & \left|\sin\left(\phi\right)\right| & -\cos\left(\phi\right)
\end{array} \right],$$
which is obtained by using only $\mu_{1}$ through $\mu_{5}$.
The top eigenvalue of $H_{t}$ can be expressed as follows:
$$T_{c}(\phi)=\left\{ \begin{array}{ll}
2\cos\left(\frac{\phi}{3}+\frac{2\pi}{3}\right)  &
\mbox{for $\scriptstyle{
-\frac{3}{2}+3j \leq \frac{\phi}{2\pi} \leq -\frac{1}{2}+3j}$}  \\
2\cos\left(\frac{\phi}{3}\right) &
\mbox{for $\scriptstyle{
-\frac{1}{2}+3j \leq \frac{\phi}{2\pi} \leq \frac{1}{2}+3j}$}  \\
2\cos\left(\frac{\phi}{3}-\frac{2\pi}{3}\right)  &
\mbox{for $\scriptstyle{
\frac{1}{2}+3j \leq \frac{\phi}{2\pi} \leq \frac{3}{2}+3j}$}
\end{array} \right. . $$

In Fig.~1, we plot the superconducting transition temperature,
$\Delta T_{c}(\Phi)\equiv T_{c}(0)-T_{c}(\Phi)=2-T_{c}(\Phi)$,
respectively, of a square loop, a hexagon loop,
and a triangle loop for $-2 \leq \Phi \leq 2$.
It is evident that these phase diagrams are qualitatively identical.
Also, the $\Delta T_{c}(\Phi)$ shown are periodic functions of 
$\Phi$ and the period of the oscillation in the flux is equal to
$\Phi_{0}$.  As expected, $\Delta T_{c}(\Phi)$ have their minima 
at $\Phi=j\Phi_{0}$ and their maxima at $\Phi=j\Phi_{0}/2$.

It is interesting to note that $\Delta T_{c}(\Phi)$ has the largest
magnitude for the triangular loop and the smallest for the hexagonal loop.
It will be seen in Sec.~X that this one-loop general behavior 
carries over to the network cases, in spite of the distinctive 
differences in the fine structure of their phase boundaries.
These results are consistent with the ones obtained 
numerically in Ref.~2.

\section{square lattice}

For the square lattice, we denote the lattice path integrals by $s_{2l}$.
In other words, $s_{2l}$ is the exact sum of the phase factors
of all $2l$-step closed paths on the square lattice.
Below $\phi/2\pi$ corresponds to the magnetic flux through an
elementary square plaquette, i.e.,
$$\frac{\phi}{2\pi}=c^2B.$$
Throughout this paper, $c$ denotes the lattice constant
of all the lattices considered in this work.
The results for $s_{2}, s_{4},\ldots, s_{12}$ are:
\begin{eqnarray*}
s_{2}&=& 4, \\
s_{4}&=& 28 + 8\cos\phi, \\
s_{6}&=& 232+144\cos\phi+24\cos2\phi, \\
s_{8}&=& 2156+2016\cos\phi+616\cos2\phi+96\cos3\phi \\
& & +16\cos4\phi, \\
s_{10}&=& 21944+26320\cos\phi+11080\cos2\phi+3120\cos3\phi \\
& & +840\cos4\phi+160\cos5\phi+40\cos6\phi,  \\
s_{12}&=&240280+ 337560\cos\phi+174384\cos2\phi  \\
& & +67256\cos3\phi+23928\cos4\phi+7272\cos5\phi \\
& & +2400\cos6\phi+528\cos7\phi+144\cos8\phi+24\cos9\phi.
\end{eqnarray*}
We have computed the lattice path integrals for the square lattice
up to $s_{138}$, which are obtained by {\it exactly summing up\/}
$\sim 10^{81}$ closed paths.  The first few lattice 
path-integrals can be quickly obtained analytically by hand. 
We have used  
Maple symbolic manipulation software to obtain lattice path integrals 
of longer lengths. For these, 
it is convenient to optimize the algorithm by exploiting 
the symmetries of the problem. These calculated lattice path integrals 
$s_{2l}$'s have enabled us to obtain the phase boundary 
up to $T_{c}^{(70)}(\phi)$.


It is instructive to explain how the first few lattice path integrals
are obtained.  This will also clarify their physical meaning. 
Since there is no path of one step for returning 
an electron to its initial site, $s_1$ is always
equal to zero.  Indeed, all lattice path integrals $s_{2l+1}$
involving an odd number of steps are equal to zero.  
Now let us compute the next lattice path integral, with two-steps.
There are four closed paths of two steps each 
[retracing each other on one bond 
($\cdot\mbox{\boldmath $\leftrightarrow$}$), 
where the dot $\cdot$ indicates the initial site], thus 
$$s_2\ =\ 4\,\cdot\!\mbox{\boldmath $\leftrightarrow$}\ 
=4\ e^{i0\phi}\ =\ 4=\ z,$$ where $z$ is the 
coordination number of the lattice.

There are $28$ closed paths of four steps each: 
four retracing twice on one bond 
($\cdot\mbox{\boldmath $\stackrel{{\textstyle \leftrightarrow}}{\leftrightarrow}$}$); 
twelve starting from a site connecting two adjacent bonds 
and retracing once on each bond 
($\mbox{\boldmath $\leftrightarrow$}\!\cdot\!\mbox{\boldmath $\leftrightarrow$}$); 
and twelve moving two bonds away and then two bonds back to the 
original site ($\cdot\mbox{\boldmath $\stackrel{{\textstyle \leftarrow}} 
{\rightarrow}$}
\mbox{\boldmath $\stackrel{{\textstyle \leftarrow}}
{\rightarrow}$}$). Since all of them enclose no area 
(i.e., no flux), then 
$$s_{4}^{\rm no \ flux}\ =\ 4\,\cdot\!
\mbox{\boldmath $\stackrel{{\textstyle \leftrightarrow}}
{\leftrightarrow}$}\,+\,
12\,\mbox{\boldmath $\leftrightarrow$}\!\cdot\!
\mbox{\boldmath $\leftrightarrow$}\,+\,
12\,\cdot\!\mbox{\boldmath $\stackrel{{\textstyle \leftarrow}}
{\rightarrow}$}
\mbox{\boldmath $\stackrel{{\textstyle \leftarrow}}
{\rightarrow}$}\ =\ 28.$$ 

Among the $4$-step closed paths, 
$8$ of them enclose adjacent square cells 
($4$ counterclockwise and $4$ clockwise) contributing 
$$\ 4\,e^{i\phi}+4\,e^{-i\phi}=8\cos\phi \ $$ 
to $\ s_4.$   
Thus it follows that 
$\ s_{4}\,=\,28\,+\,8\,\cos\phi$. Higher-order 
integrals $s_{2l}$ can be similarly constructed.


It is straightforward to compute the non-zero parameters $b_{n}$ from
the obtained results for $s_{2l}$. The corresponding truncated
Hamiltonians, $H^{(n)}$, can then be readily constructed.
For instance, the second-order
truncation of the Hamiltonian is
$$H^{(2)}=\left [ \begin{array}{cc}
0 & 2 \\
2 & 0
\end{array} \right].$$
Its corresponding top eigenvalue is $T^{(2)}_{c}(\phi)=2$, which does not
depend on $\phi$. This is understandable from the fact that the shortest
length for a closed path on the square lattice to enclose the magnetic flux
is for $l=4$ while $H^{(2)}$ only contains elements derived from $\mu_{2}$.
The third-order truncation of the Hamiltonian is
$$ H^{(3)}= \left [ \begin{array}{ccc}
0 & 2 & 0  \\
2 & 0 & \sqrt{3+2\cos\phi}  \\
0 & \sqrt{3+2\cos\phi} & 0   \end{array} \right ].$$
Its corresponding top eigenvalue is
$$T^{(3)}_{c}(\phi)=\sqrt{7+2\cos\phi} \ .$$
The fourth-order truncation of the Hamiltonian,  $H^{(4)}$, is
$$\left [ \begin{array}{cccc}
0 & 2 & 0 & 0  \\
2 & 0 & \scriptstyle{\sqrt{3+2\cos\phi}}  & 0  \\
0 & \scriptstyle{\sqrt{3+2\cos\phi}} & 0 &
\sqrt{\frac{3+8\cos\phi+8\cos^{2}\phi}{3+2\cos\phi}}  \\
0 & 0 & \sqrt{\frac{3+8\cos\phi+8\cos^{2}\phi}{3+2\cos\phi}}
& 0    \end{array} \right ]. $$
Its corresponding top eigenvalue is
$$T^{(4)}_{c}(\phi)=\sqrt{2}
\sqrt{\frac{3\cos^{2}\phi+7\cos\phi+6+\alpha}{3+2\cos\phi}},$$
where
$$\alpha=\sqrt{9\cos^{4}\phi+26\cos^{3}\phi+45\cos^{2}\phi+54\cos\phi+27}.$$

In Fig.~2, we show the superconducting transition temperatures,
$$\Delta T^{(n)}_{c}(\Phi)=T_{c}(0)-T_{c}^{(n)}(\Phi),$$ 
as functions of $\Phi\equiv \phi/2\pi$ for various values of $n$, 
for the square network obtained from the truncated Hamiltonians $H^{(n)}$.
Here $T_{c}(0)$ equals $4$, which is the largest eigenvalue of
tight-binding electrons confined on the square lattice in the absence
of a magnetic field.
It is important to stress that as the order of approximation is increased, more
geometrical information of the lattice is included in the interference
treatment, and more fine structures are resolved. 
At every step, i.e., for a given size of the network, we can observe 
the corresponding dips appearing and then becoming sharper.
We emphasize that our highest-order approximant, 
$T_{c}^{(70)}(\Phi)$ has closely reached the infinite system 
size limit, $\Delta T_{c}(\Phi)$. The flux values where
the cusps/dips occur have also been labeled.

\section{self-similarity in the phase boundary of the superconducting
square wire network}

In this section, we explicitly demonstrate an important
property: the self-similarity of the phase boundary of 
the superconducting square wire network.
This is exemplified in Fig.~3, where we use
$\Delta T^{(70)}_{c}(\Phi)$ for $\Delta T_{c}(\Phi)$ and omit the
superscript. In (a), we plot $\Delta T_{c}(\Phi)$ for
$\Phi$ in the interval between $0$ and $1$. In (b) and (c),
we plot $\Delta T_{c}(\Phi)$ for $\Phi$, respectively, in the ranges
$[0.333 \simeq 1/3, \ 0.4765]$  and  $[0.5235,\ 0.667 \simeq 2/3]$.
Figures (b) and (c) can be regarded as the first generation of the
original diagram (a), in the sense that (b) is enlarged
from the maximum in the left part of (a) and (c) is enlarged
from the maximum in the right part of (a). 

This enlargement process is continued as follows:
(d) with $\Phi \in [0.375=3/8,\ 0.3978]$ and
(e) with $\Phi \in [0.4025,\ 0.4286\simeq 3/7]$ are, respectively,
the enlargements of the left and right maxima of (b). Similarly,
(f) with $\Phi \in [0.5714 \simeq 4/7,\ 0.5975]$ and
(g) with $\Phi \in [0.6022,\ 0.625=5/8]$ are, respectively,
the enlargements of the left and right maxima of (c).
Figures (d), (e), (f), and (g) can be regarded as the second generation
of the original phase diagram (a). In this way,
it is straightforward to deduce
that the third generation of (a) will consist of $8$ phase diagrams:
each of (d), (e), (f), and (g) contributes two diagrams.
It is evident that these phase diagrams resemble one another except that
the phase diagrams gradually become asymmetric.

As shown in these figures, we also label the values of $\Phi$
indicating the cusps/dips in $\Delta T_{c}(\Phi)$. These nine flux values are
characteristic of each phase diagram.
Indeed, there are general relations between these sets of flux values
in different generations.
Let $\{p_{0}/q_{0}\}$ represent the set of these flux values in (a), i.e.,
$p_{0}/q_{0}=1/4$, $2/7$, $1/3$, $2/5$, $1/2$, $3/5$, $2/3$, $5/7$,
and $3/4$. Denoting the set of characteristic flux values in any of the
phase diagrams in the first generation by $\{p_{1}/q_{1}\}$,
we find that the corresponding flux values in (b) are given by
$$ \frac{p_{1}}{q_{1}} = \frac{ q_{0} }{ 3q_{0}-p_{0} } \ ,$$  
and those in (c) are given by 
$$ \frac{ p_{1} }{ q_{1} }= \frac{ p_{0}+q_{0} }{ p_{0}+2q_{0} }. $$
For instance, given $p_{0}/q_{0}=1/2$ in (a), we have the corresponding
$$ p_{1}/q_{1}=2/(6-1)=2/5 $$ 
in (b) and 
$$ p_{1}/q_{1}=(1+2)/(1+4)=3/5 $$ 
in (c).  Furthermore, let $\{p_{2}/q_{2}\}$ stand for 
the sets of the corresponding flux values in the 
second-generation diagrams.  In the second-generation diagrams
[(d)--(g)] only $5$ characteristic cusps/dips out of 
$9$ are observable.  There we find that the
$p_{2}/q_{2}$ in (d) are related to the $p_{1}/q_{1}$ in 
(b) by 
$$ \frac{ p_{2} }{ q_{2} } \ =\ \frac{ q_{1} }{ 3q_{1}-p_{1} },$$
$p_{2}/q_{2}$ in (e) are related to $p_{1}/q_{1}$ in (c) by
$p_{2}/q_{2}=q_{1}/(3q_{1}-p_{1}),$
$p_{2}/q_{2}$ in (f) are related to $p_{1}/q_{1}$ in (b) by
$$ \frac{ p_{2} }{ q_{2} } \ =\ \frac{ p_{1}+q_{1} }{ p_{1}+2q_{1}},$$ 
and $p_{2}/q_{2}$ in (g) are related to $p_{1}/q_{1}$ in (c) by
$p_{2}/q_{2}=(p_{1}+q_{1})/(p_{1}+2q_{1}).$

We now summarize our construction of the hierarchy of 
these phase diagrams.
As discussed previously, every diagram can generate two diagrams
to the next generation: one is enlarged from the left maximum 
and the other from the right maximum of this diagram.
Thus, starting from the original phase diagram, i.e., $\Delta T_{c}(\Phi)$
for $\Phi \in [0,1]$, we can generate $2^{n}$ diagrams to the $n$th
generation for $n \geq 1$.
Furthermore, each diagram covers a distinct range of $\Phi$
from $\Phi_{{\rm min}}$ to $\Phi_{{\rm max}}$.
Let us arrange these diagrams in the following way, as we did in
Fig.~3.
We put all the diagrams belonging to the same generation in a row
in such an order that from the left to the right $\Phi_{{\rm min}}$
(or $\Phi_{{\rm max}}$) increases from the smallest to the largest.
It is evident that half of them ($2^{n-1}$ diagrams) have
$\Phi_{{\rm max}} < 1/2$ and the other half have $\Phi_{{\rm min}} > 1/2$.
It is not difficult to see that this kind of arrangement
will be automatically satisfied in the following way.
Following the same order of the diagrams in the previous generation and
using them one by one, we put two new generated diagrams side by side
with the one from the left maximum to the left and the one from the right
maximum to the right.
It is interesting to notice that, for each generation, the diagrams
located at the left part of $\Phi=1/2$ are mirror images of those located
at the right part. This symmetry originates from the property that the phase
diagram of $\Delta T_{c}(\Phi)$ with $\Phi \in [0,1]$ is symmetric around
$\Phi=1/2$.

Indeed, there are one-to-one correspondences between the sets of the
characteristic flux values, where cusps/dips in the phase boundaries
occur, in different generations.
Let us label the diagrams from left to right in the $n$th generation
by ${\cal D}^{(n)}_{i}$ with $i$ running
from 1 to $2^{n}$.
Similarly, the diagrams in the $(n+1)$th generation are labeled by
${\cal D}^{(n+1)}_{i}$ with $i$ running from 1 to $2^{n+1}$.
Now let $\{p_{n}/q_{n}\}$ represent the sets of the flux values
characterizing the cusps/dips in $\Delta T_{c}(\Phi)$ in any of the phase
diagrams in the $n$th generation
and $\{p_{n+1}/q_{n+1}\}$ be the sets belonging to the diagrams in the
$(n+1)$th generation.
The relations between the $(p_{n+1}/q_{n+1})$'s and the
$(p_{n}/q_{n})$'s are as follows.
For $1 \leq i \leq 2^n$, the $p_{n+1}/q_{n+1}$ in the diagram
${\cal D}^{(n+1)}_{i}$  [one of the diagrams in
the $(n+1)$th generation that located on the left hand side of
$\Phi=1/2$] is related to the $p_{n}/q_{n}$ in
${\cal D}^{(n)}_{i}$ by
$$\frac{p_{n+1}}{q_{n+1}}=\frac{q_{n}}{3q_{n}-p_{n}}$$
and for $2^n+1 \leq i \leq 2^{n+1}$, the $p_{n+1}/q_{n+1}$ in the diagram
${\cal D}^{(n+1)}_{i}$  [the second half of the diagrams in
the $(n+1)$th generation that located on the right hand side of
$\Phi=1/2$] is related to the $p_{n}/q_{n}$ in
${\cal D}^{(n)}_{i-2^n}$ by
$$\frac{p_{n+1}}{q_{n+1}}=\frac{p_{n}+q_{n}}{p_{n}+2q_{n}}.$$

Self-similarity in the $\Delta T_{c}(\Phi)$ curve is a consequence
of the fractal energy spectrum of Bloch electrons in a magnetic 
field which was examined in detail by Hofstadter.\cite{18} 
However,  as far as we are aware, the explicit derivation 
of the self-similarity of the measurable part, the lowest 
energy state, was not presented before.

Recently, the influence of classical chaos on this so 
called ``Hofstadter's butterfly'' has been studied.\cite{19} 
Furthermore, a semiclassical theory for the dynamics of
electrons in a magnetic Bloch band has been developed and used 
to explain the clustering structure of the spectrum.\cite{20}

\section{honeycomb lattice}

For the honeycomb lattice, we denote the lattice path integrals by $h_{2l}$.
In other words, $h_{2l}$ is the exact sum of the phase factors
of all $2l$-step closed paths on the honeycomb lattice.
In this section, $\phi/2\pi$ corresponds to the magnetic flux through an
elementary honeycomb plaquette, i.e.,
$$\frac{\phi}{2\pi}=\frac{3\sqrt{3}c^2B}{2\Phi_{0}}.$$
The results for $h_{2}, h_{4},\ldots, h_{20}$ are:
\begin{eqnarray*}
h_{2}&=& 3, \\
h_{4}&=& 15, \\
h_{6}&=&87+6\cos\phi,  \\
h_{8}&=&543+96\cos\phi,  \\
h_{10}&=&3543+ 1080\cos\phi+30\cos2\phi, \\
h_{12}&=&23859+10560\cos\phi+726\cos2\phi+ 24\cos3\phi, \\
h_{14}&=&164769+96096\cos\phi+11130\cos2\phi+ 798\cos3\phi \\
& & + 42\cos4\phi, \\
h_{16}&=&1162719+839040\cos\phi+138720\cos2\phi \\
& &+15648\cos3\phi +1536\cos4\phi+96\cos5\phi, \\
h_{18}&=&8363895+7143210\cos\phi+1537668\cos2\phi \\
& &+237714\cos3\phi+33246\cos4\phi+3834\cos5\phi \\
& &+252\cos6\phi+18\cos7\phi, \\
h_{20}&=&61216275+59862000\cos\phi+15829200\cos2\phi \\
& & +3103320\cos3\phi+555390\cos4\phi+89520\cos5\phi  \\
& & +10920\cos6\phi+ 1320\cos7\phi+120\cos8\phi.
\end{eqnarray*}

Notice that $h_2$ and $h_4$ involve paths that enclose 
zero net flux.  There are three closed paths of 2-steps each. 
Thus, $h_2=3$, the coordination number of the lattice.
$h_6$ is the first lattice path integral 
with a net flux (in this case flux through one hexagon). 
There are three counter-clockwise and three clockwise 
six-step paths going through a hexagon.  Thus, the 
term $6\cos\phi$ in $h_6$. It is possible to derive 
the first few path-integrals analytically ``by hand", 
by just counting paths and keeping track of the enclosed 
flux.  The longer-length ones can be computed via symbolic 
manipulation software.

We have computed the lattice path integrals for the honeycomb lattice
up to $h_{206}$, which are obtained by {\it exactly\/} summing up
$\sim 10^{96}$ closed paths. These calculated $h_{2l}$'s have enabled us
to obtain the phase boundary up to $T_{c}^{(104)}(\phi)$.

It is straightforward to compute the non-zero parameters $b_{n}$ from
the obtained results for $h_{2l}$. The corresponding truncated
Hamiltonians, $H^{(n)}$, can then be readily constructed.
For instance, the second-order
truncation of the Hamiltonian is
$$H^{(2)}=\left [ \begin{array}{cc}
0 & \sqrt{3} \\
\sqrt{3} & 0
\end{array} \right].$$
Its corresponding top eigenvalue is $T^{(2)}_{c}=\sqrt{3}$.
The third-order truncation of the Hamiltonian is
$$ H^{(3)}= \left [ \begin{array}{ccc}
0 & \sqrt{3} & 0  \\
\sqrt{3} & 0 & \sqrt{2} \\
0 & \sqrt{2} & 0  \end{array} \right ].$$
Its corresponding top eigenvalue is $T^{(3)}_{c}=\sqrt{5}$.
Both $T^{(2)}_{c}$ and $T^{(3)}_{c}$ are independent of $\phi$.
This is understandable from the fact that the shortest
length for a closed path on the honeycomb lattice to enclose the magnetic flux
is for $l=6$ while $H^{(2)}$ and $H^{(3)}$ only contain elements derived
from $\mu_{2}$ and $\mu_{4}$.
The fourth-order truncation of the Hamiltonian is
$$ H^{(4)}= \left [ \begin{array}{cccc}
0 & \sqrt{3} & 0 & 0  \\
\sqrt{3} & 0 & \sqrt{2} & 0  \\
0 & \sqrt{2} & 0 & \sqrt{2+\cos\phi} \\
0  & 0 & \sqrt{2+\cos\phi} & 0 \end{array} \right ].$$
Its corresponding top eigenvalue is
$$T^{(4)}_{c}(\phi)=\frac{1}{2}\sqrt{14+2\cos\phi
+2\sqrt{25+2\cos\phi+\cos^{2}\phi}}.$$

In Fig.~4, we show the superconducting transition temperatures,
$\Delta T^{(n)}_{c}(\Phi)=T_{c}(0)-T_{c}^{(n)}(\Phi)$, as
functions of $\Phi\equiv \phi/2\pi$ for various $n$ for the
honeycomb network obtained from the truncated Hamiltonians $H^{(n)}$.
Here $T_{c}(0)$ equals $3$, which is the largest eigenvalue of
tight-binding electrons confined on the honeycomb lattice in the absence
of a magnetic field.

We observe that as the order of approximation is increased, 
more geometrical information of the lattice is included 
in the {\it interference treatment}, and more fine structures are resolved.  
This explains {\it the origin of the fine structure observed}: the more 
geometric information on the lattice is explored by the paths of 
the electrons, the sharper the fine structures.

We emphasize that our
highest-order approximant, $T_{c}^{(104)}(\Phi)$ has closely reached the
infinite system size limit, $\Delta T_{c}(\Phi)$. The flux values where
the cusps/dips occurred have also been labeled. In general, besides the cusp
at $\Phi=1/2$, there are cusps at 
$$ \Phi \ =\ \frac{m}{2m+1} $$ and 
$$\Phi\ =\ \frac{m+1}{2m+1}$$
with $m \geq 1$.
Our computed phase boundary compares well with the observed cusps 
present in experiments \cite{nec1,nec2}.  

\section{triangular lattice}

For the triangular lattice, we denote the lattice path integrals by $t_{l}$.
In other words, $t_{l}$ is the exact sum of the phase factors
of all $l$-step closed paths on the triangular lattice.
In this section, $\phi/2\pi$ corresponds to the magnetic flux through an
elementary triangular plaquette, i.e.,
$$\frac{\phi}{2\pi}=\frac{\sqrt{3}c^2B}{4\Phi_{0}}.$$
The results for $t_{2}$ through $t_{10}$ are:
\begin{eqnarray*}
t_{2}&=& 6, \\
t_{3}&=& 12\cos\phi, \\
t_{4}&=&66+24\cos2\phi,  \\
t_{5}&=&300\cos\phi+60\cos3\phi, \\
t_{6}&=&1020+840\cos2\phi+168\cos4\phi+12\cos6\phi,  \\
t_{7}&=&6888\cos\phi+2604\cos3\phi+504\cos5\phi+84\cos7\phi, \\
t_{8}&=&19890+23904\cos2\phi+8568\cos4\phi+1968\cos6\phi \\
& & +432\cos8\phi+48\cos10\phi, \\
t_{9}&=&164124\cos\phi+85944\cos3\phi+29628\cos5\phi \\
& & +8496\cos7\phi+1980\cos9\phi+432\cos11\phi \\
& & +36\cos13\phi, \\
t_{10}&=&449976+654840\cos2\phi+317940\cos4\phi  \\
& & +114360\cos6\phi+37560\cos8\phi+10380\cos10\phi  \\
& & +2700\cos12\phi+540\cos14\phi+60\cos16\phi.
\end{eqnarray*}


Here we explain how the first few lattice path integrals
are obtained.  
Since there is no path of one step for returning 
an electron to its initial site, $t_1$ is always
equal to zero. There are six closed paths of two 
steps each [retracing each other on one bond 
($\cdot\mbox{\boldmath $\leftrightarrow$}$), 
where the dot $\cdot$ indicates the initial site], thus 
$$t_2\ =\ 6\,\cdot\!\mbox{\boldmath $\leftrightarrow$}\ 
=6\ e^{i0\phi}\ =\ 6=\ z,$$ where $z$ is the 
coordination number of the lattice. 

There are 12 three-step closed-paths enclosing a 
triangular cell [three counterclockwise 
($\cdot\!\!\stackrel{\leftarrow}{\mbox{\boldmath $\bigtriangledown$}}$), 
and three clockwise 
($\cdot\!\!\stackrel{\rightarrow}{\mbox{\boldmath $\bigtriangledown$}}$)]. 
Thus
$$t_{3}\ =\ 
6\,\cdot\!\!\stackrel{\leftarrow}{\mbox{\boldmath $\bigtriangledown$}}\,+\,
6\,\cdot\!\!\stackrel{\rightarrow}{\mbox{\boldmath $\bigtriangledown$}}\ 
=\ 6\,e^{i\phi}\,+\,6\,e^{-i\phi}\ =\ 12\,\cos\phi.$$ 

There are $66$ closed paths of four steps each
enclosing zero flux each: six retracing twice on one bond 
($\cdot\mbox{\boldmath $\stackrel{{\textstyle \leftrightarrow}}{\leftrightarrow}$}$); 
thirty starting from a site connecting two adjacent bonds 
and retracing once on each bond 
($\mbox{\boldmath $\leftrightarrow$}\!\cdot\!\mbox{\boldmath $\leftrightarrow$}$); 
and thirty moving two bonds away and then 
two bonds back to the original site 
($\cdot\mbox{\boldmath $\stackrel{{\textstyle \leftarrow}}
{\rightarrow}$}
\mbox{\boldmath $\stackrel{{\textstyle \leftarrow}}
{\rightarrow}$}$). Since all of them enclose no area 
(i.e., no flux), then 
$$t_{4}^{\rm no \ flux}\ =\ 6\,
\cdot\!\mbox{\boldmath $\stackrel{{\textstyle \leftrightarrow}}{\leftrightarrow}$}\,+\,
30\,\mbox{\boldmath $\leftrightarrow$}\!\cdot\!
\mbox{\boldmath $\leftrightarrow$}\,+\,
30\,\cdot\!\mbox{\boldmath $\stackrel{{\textstyle \leftarrow}}
{\rightarrow}$}
\mbox{\boldmath $\stackrel{{\textstyle \leftarrow}}
{\rightarrow}$}\ =\ 66.$$

Among the $4$-step closed paths, $24$ 
of them enclose adjacent cells enclosing two triangles 
($12$ counterclockwise and $12$ clockwise) and contribute 
$$ t_4^{\rm two \ cells \ only} = 12 e^{2 i\phi}+12 e^{-2 i\phi} = 24 \cos2\phi$$ 
to $t_4.$   
Thus, it follows that $t_{4}\,=\,66\,+\,24\,\cos2\phi.$


Note that $t_{2l}$ ($t_{2l+1}$) consist of only even (odd) harmonics of the
flux. We have computed the lattice path integrals for the triangular lattice
up to $t_{119}$, which are obtained by exactly summing up
$\sim 10^{90}$ closed paths. These calculated $t_{l}$'s have enabled us
to obtain the phase boundary up to $T_{c}^{(60)}(\phi)$.

By using the calculated results for $t_{l}$, the parameters $a_{n}$ and
$b_{n}$, and subsequently the corresponding truncated
Hamiltonians, $H^{(n)}$, can be obtained.
For instance, the second-order truncation of the Hamiltonian is
$$H^{(2)} =\left[ \begin{array}{cc}
0 & \sqrt{6} \\
\sqrt{6}  & 2\cos\phi
\end{array} \right].$$
Its corresponding top eigenvalue is
$$T^{(2)}_{c}(\phi)=\cos\phi+\sqrt{6+\cos^{2}\phi}.$$
The third-order truncation of the Hamiltonian is
$$ H^{(3)}= \left[ \begin{array}{ccc}
0 & \sqrt{6} & 0  \\
\sqrt{6} & 2\cos\phi & \sqrt{1+4\cos^{2}\phi} \\
0 & \sqrt{1+4\cos^{2}\phi} &
{\displaystyle \frac{-8\cos\phi+16\cos^{3}\phi}{1+4\cos^{2}\phi}}
\end{array} \right].$$
Its corresponding top eigenvalue $T^{(3)}_{c}(\phi)$ can also be obtained
analytically.

In Fig.~5, we show the superconducting transition temperatures,
$\Delta T^{(n)}_{c}(\Phi)=T_{c}(0)-T_{c}^{(n)}(\Phi)$, as
functions of $\Phi\equiv \phi/2\pi$ for various $n$ for the
triangular network obtained from the truncated Hamiltonians $H^{(n)}$.
Here $T_{c}(0)$ equals $6$, which is the largest eigenvalue of
tight-binding electrons confined on the triangular 
lattice in the absence of a magnetic field.
The following physical picture is clear from those plots:
as the order of approximation is increased, more geometrical 
information of the lattice is included in the interference
treatment, and more fine structures are resolved. 

Our highest-order approximant, $T_{c}^{(60)}(\Phi)$ has closely reached the
infinite system size limit, $\Delta T_{c}(\Phi)$. The flux values where
the cusps occurred have also been labeled. In general, besides the cusps 
at $\Phi=1/2$, $1/5$, $4/5$, $5/16$, and $11/16$,
there are cusps/dips at 
$$\Phi=\frac{m}{2m+2}$$ and 
$$\Phi=\frac{m+2}{2m+2}$$
with $m \geq 1$.

\section{kagome lattice}

Our computed phase boundary compares well with the observed cusps 
present in a series of interesting experiments \cite{nec1,nec2}.  

For the kagome lattice,\cite{lin-kagome,nec1,nec2,huse,premi,fabiana}
we denote the lattice path integrals by $k_{l}$.
In other words, $k_{l}$ is the exact sum of the phase factors
of all $l$-step closed paths on the kagome lattice.
Here $\phi/2\pi$ corresponds to the magnetic flux through an
elementary triangular plaquette, i.e.,
$$\frac{\phi}{2\pi}=\frac{\sqrt{3}c^2B}{4\Phi_{0}}.$$
The results for $k_{2}$ through $k_{11}$ are:
\begin{eqnarray*}
k_{2}&=& 4, \\
k_{3}&=& 4\cos\phi ,\\
k_{4}&=& 28,\\
k_{5}&=& 60\cos\phi , \\
k_{6}&=&244+16\cos2\phi+4\cos6\phi , \\
k_{7}&=&756 \cos\phi +28\cos7\phi, \\
k_{8}&=&2412+416\cos2\phi+96\cos6\phi+80\cos8\phi, \\
k_{9}& = &9216\cos\phi+76\cos3\phi+36\cos5\phi   \\
& & +756\cos7\phi+120\cos9\phi,    \\
k_{10}& = &25804+7560\cos2\phi+1860\cos6\phi  \\
& & +2480\cos8\phi+100\cos10\phi+20\cos14\phi,  \\
k_{11}& = &112420\cos\phi+2816\cos3\phi+1276\cos5\phi \\
& & +14608\cos7\phi+4400\cos9\phi+44\cos11\phi   \\
& & + 44\cos13\phi+176\cos15\phi. \\
\end{eqnarray*}
Note that $k_{2l}$ ($k_{2l+1}$) comprise only even (odd) harmonics of the
flux. We have computed the lattice path integrals for the kagome lattice
up to $t_{99}$, which are obtained by exactly summing up $\sim 10^{58}$
closed paths. These calculated $k_{l}$'s have enabled us
to obtain the phase boundary up to $T_{c}^{(50)}(\phi)$.

By using the calculated results for $k_{l}$, the parameters $a_{n}$ and
$b_{n}$, and subsequently the corresponding truncated
Hamiltonians, $H^{(n)}$, can be obtained.
For instance, the second-order truncation of the Hamiltonian is
$$H^{(2)} =\left[ \begin{array}{cc}
0 & 2 \\
2 & \cos\phi
\end{array} \right].$$
Its corresponding top eigenvalue is
$$T^{(2)}_{c}(\phi)=\frac{1}{2}\left(\cos\phi+\sqrt{16+\cos^{2}\phi}\right).$$
The third-order truncation of the Hamiltonian is
$$ H^{(3)}= \left[ \begin{array}{ccc}
0 & 2 & 0  \\
2 & \cos\phi & \sqrt{3-\cos^{2}\phi} \\
0 & \sqrt{3-\cos^{2}\phi} &
{\displaystyle \frac{\cos\phi+\cos^{3}\phi}{3-\cos^{2}\phi}}
\end{array} \right].$$
Its corresponding top eigenvalue $T^{(3)}_{c}(\phi)$ can also be obtained
analytically.

In Fig.~6, we show the superconducting transition temperatures,
$\Delta T^{(n)}_{c}(\Phi)=T_{c}(0)-T_{c}^{(n)}(\Phi)$, as
functions of $\Phi\equiv \phi/2\pi$ for various $n$ for the
kagome network obtained from the truncated Hamiltonians $H^{(n)}$.
Here $T_{c}(0)$ equals $4$ which is the largest eigenvalue of
tight-binding electrons confined on the kagome lattice in the absence
of the magnetic field.
It is seen that as the order of the approximation is increased, more
geometrical information of the lattice is included in the interference
treatment, and more fine structures are resolved. We emphasize that our
highest-order approximant, $T_{c}^{(50)}(\Phi)$ has closely reached the
infinite system size limit, $\Delta T_{c}(\Phi)$. The flux values where
the cusps/dips occurred have also been labeled.
Our computed phase boundary compares well with the observed cusps 
present in a series of interesting experiments \cite{nec1,nec2}.  
See also the systematic calculations in Ref.~\cite{huse}.

\section{discussion}

In the following, we discuss the general trends in the approximants
for these phase diagrams presented in the above sections.

\subsection{Comparison of the structure in the phase boundaries}

In the lower-order approximants, the first noticeable development in the
phase boundaries of square, honeycomb, and triangular lattice is the
formation of dips when the flux per elementary plaquette is equal to
$m\Phi_{0}/2$, where $m$ is an integer.
When the order of approximation is increased, the dips at
$\Phi=1/2$ become sharper and at the same time more fine structures
(other local minima) begin to emerge. Eventually, the dips at various
different flux values become cusps.

It is interesting to notice that, among these three lattices, the
development of the cusps is most rapid for the triangular case while the
honeycomb is slowest. This difference originates from the fact that for
identical length, lattice path integrals for the triangular lattice
contain the richest {\it quantum interference\/} effects because 
the number of paths and the areas they enclose are both the largest.
For the kagome network, the rapid development of cusps at
$$\Phi^{{\rm kagome}} = \frac{1}{8},\ \frac{1}{4}, \ \frac{3}{8} ,\ 
\frac{5}{8} ,\ \frac{3}{4} ,\ {\rm and} \ \frac{7}{8} $$
\noindent
can be seen from lower-order approximants. For additional 
discussion on the kagome case, see Ref.~\cite{lin-kagome}. 
For extensions of these techniques to other problems, see Ref.~\cite{lin}.

In general, the resulting phase diagrams---with the occurrence 
of cusps/dips at different sets of flux values---are a direct 
consequence of the geometries of the lattices, which is explicitly 
reflected in the corresponding expressions of the lattice path integrals.
We stress that our evaluation of the lattice path integrals to extremely
long lengths has enabled our calculated $T_{c}(B)$ to achieve
close convergence to the infinite system size. Indeed, for $n\simeq 10$,
important features in the phase boundaries of square, triangular,
and kagome networks are well developed.

Finally, in order to facilitate a comparison between the different 
phase boundaries, in Fig.~7 we plot
$\Delta T_{c}(\Phi)$ as a function of $\Phi$ for the square, 
honeycomb, triangular, and kagome superconducting networks.
Here the $\Delta T_{c}(\Phi)$'s are taken from their respective 
highest-order approximants and $\Phi$ is the flux through their 
respective elementary cells as discussed in the previous sections.
Here we omit the subscripts indicating the order of approximation.
The values of the magnetic flux corresponding to a number of
prominent cusps/dips are also labeled.
%

\subsection{Comparison of the phase boundaries of the single-loop
and lattice cases}

From Figs.~1 and 7 (a-c), we can readily see the differences between the phase
boundaries of a single superconducting cell and its corresponding
superconducting network.
For both cases, $\Delta T_{c}$ vary periodically with the magnetic
flux through a single elementary cell and have the same period $\Phi_{0}$ 
of oscillation.  We now focus on $\Delta T_{c}(\Phi)$ for $\Phi$
in the interval between $0$ and $1$. $\Delta T_{c}(\Phi)$ is
symmetric around $\Phi=1/2$. However, there are many distinct features
between $\Delta T_{c}$ of a single cell and that of a network. These
differences are due to long-range correlations of the many-loops effect 
present in the lattice.
For a single superconducting cell, $\Delta T_{c}(\Phi)$ increases
monotonically from $\Phi=0$ to $\Phi=1/2$ and decreases monotonically
from $\Phi=1/2$ to $\Phi=1.$ 
The maximum at $\Phi=1/2$ exhibits a sharp peak. Indeed, the
overall shape of $\Delta T_{c}(\Phi)$ resembles the combination of two
identical half parabolas both reaching their maximum at $\Phi=1/2$.
On the contrary, the overall shape of $\Delta T_{c}(\Phi)$ for
the corresponding superconducting networks look like downward parabolas 
with many local cusps between $\Phi=0$ and $\Phi=1$.
The most prominent cusps are located at $\Phi=1/2$. The positions of other
cusps/dips depend on the underlying lattice types of the networks.

\subsection{Differences between our approach and the traditional
moments and Lanczos methods}

In electronic structure calculations there is a method to compute the
density of states called the moments method.  This is similar to our
approach in the sense that $\mu_{l}$ can be interpreted as the
moments, $\langle \Psi_i |H^{l}| \Psi_i \rangle$.
However there are several important differences between the
standard ``moments method''
and our problem.  The typical use of the moments method:
(i) focuses on the computation of electronic density of
states (instead of superconducting $T_c$'s);
(ii) is totally numerical (instead of mostly analytical);
(iii) is done at zero magnetic field (instead of obtaining
expressions with an explicit field dependence);
(iv) does not focus on the explicit computation of lattice
path integrals; and
(v) does not study the physical effects of quantum interference
(which is at the heart of our calculation and physical interpretation).
In conclusion, the traditional use of the moments method in solid state
is significantly different from the approach and problem studied here.

Another way to diagonalize Hamiltonians is called the Lanczos method.
This method directly obtains the tridiagonal form, without
computing the moments; thus differs in a significant way from the approach
used here (where the explicit computation of the moments is one of our
goals, since they can be used for other electronic property calculations).
Furthermore, it is not convenient to use standard Lanczos method in our
particular problem because it is extremely difficult to directly derive the
parameters and the states of the iterative tridiagonalization
procedure.  This is so because of the presence of the magnetic
field.  On the other hand, the moments method provides standard 
procedures to diagonalize a matrix after the moments are computed.


\subsection{Commensurability and other Matching Effects}

An essential physics issue in this problem is {\it commensurability}.
Another one is {\it quantum interference}---due to the motion of 
electrons in multiconnected geometries.  This section briefly 
overviews related systems where commensurability and matching 
effects (due to externally applied magnetic fields) play an 
important role.  The first example will be flux pinning.

Flux pinning in type II superconductors is of both technological and 
scientific interest.  While most experiments focus on the effects of 
random pinning distributions, some investigations have been carried out 
on periodic arrays of pinning sites \cite{1}.
These find striking peaks in the magnetization 
\cite{Metulshko} and critical current $ J_{c} $.
These peaks are believed to arise from the greatly enhanced pinning 
that occurs when parts of the vortex lattice become commensurate 
with (i.e., match) the underlying periodic array of pinning sites.   
Under such conditions, high-stability vortex configurations are produced 
which persist under an increasing current or external field.

Other important vortex matching effects have also recently been 
observed in a variety of different superconducting systems 
\cite{hunnekes,JJ,oussena},
including long Josephson junctions with periodically-spaced grooves \cite{JJ}, 
superconducting networks \cite{reviews}, 
and the matching of the VL to the crystal structure of 
YBa$_2$Cu$_3$O$_7$ due to intrinsic pinning\cite{oussena}.  

Matching effects between a vortex lattice and periodic pinning 
arrays produce a rich variety of effects \cite{charles}.
The dynamics observed in these systems is quite different 
from the one found for random arrays of pinning sites 
(see, e.g., Ref.~\cite{olson} 
and references therein).

Non-superconducting systems also exhibit magnetic-field-tuned matching 
effects, notably in relation to electron motion in periodic structures
where unusual behaviors arise due to the incommensurability of the
magnetic length with the lattice spacing. 
A recent example of these is provided by the anomalous Hall plateaus 
of ``electron  pinball''\cite{pinball} orbits scattering from 
a regular array of antidots.

Commensurate effects also play central roles in many other 
areas of physics, including plasmas, nonlinear dynamics\cite{NL}, 
the growth of crystal surfaces, domain walls in incommensurate solids, 
quasicrystals, Wigner crystals, as well as spin and charge density 
waves.  The next section discusses in some detail an example 
in nonlinear dynamics (which is virtually unknown in the solid 
state literature) that produces a fractal phase boundary which 
is strikingly similar to the one measured for square superconducting 
networks---because both are determined by commensurability effects.



\subsection{Kagome-pinned vortices:  "Correlated Melting" and Cooperative 
Ring Excitations for Doubly-Degenerate Ground States}

Notice that the fluxoid configurations for $f=1/2$ for the 
superconducting networks (e.g., Fig. 3 of Yi Xiao, et al, in the companion 
article\cite{nec1}) has two ground states that correspond to 
the two degenerate ground states of the second matching field of vortices in 
type II superconductors with a kagome periodic array of pinning sites.  
The latter has been systematically studied in Ref.~\cite{fabiana}.

The kagom\'e pinning potential at the second matching field shows novel and 
interesting dynamics as a function of temperature,\cite{fabiana} including a phase with 
rotating vortex triangles caged by kagom\'e hexagons (``cooperative ring 
elementary excitations"), and there is geometric frustration for 
$T \rightarrow 0$ with a doubly-degenerate ground state.
At finite temperatures, the three vortices inside the kagome' hexagon 
can move and rotate by 60 degrees.  This is done cooperatively by 
the three vortices.  They motion is similar to the 
``cooperative ring exchange" motion proposed by Feynman 
for elementary excitations in Helium 4.  

In other words, for the second matching field for the kagom\'e pinning lattice, 
the elementary excitation of the three interstitial vortices is a 
60 degree rotation, rotating as a cooperative ring.  These type of 
collective or correlated cooperative ring exchange has also 
been studied in the context of the quantum Hall effect.

For increasing temperatures, a novel type of melting \cite{fabiana}
appears. This can be described as ``correlated melting" in the 
sense that the ``triangle" or ``loop" first melts in the angular 
coordinate, while the radial coordinate does not melt until 
much higher temperatures are reached.  The elementary excitations 
are the thermal analog of certain types of {\it squeezed states\/} 
(where fluctuations strongly affect a coordinate and less the 
other coordinate).  They are also analogs of the {\it rotational 
isomers\/} or ``comformations" that are often found in molecules, 
where three atoms and molecules can cooperatively oscillate 
back and forth between two degenerate ground states.

This type of ``controlled melting" or ``correlated melting"\cite{fabiana} 
of the particles inside a potential energy trap could also be 
visualized with a colloidal suspension surrounded by six pinned 
(e.g., by laser tweezers) charged particles.  This type of 
``vortex-analog" experiment is easier to visualize 
(e.g., via optical microscope) than using 
vortices.  Still, Lorentz microscopy techniques\cite{tonomura} 
could directly image such motions in the vortex case.


\subsection{Fractal Phase boundaries and Fractal Boundaries of Basins of 
Attraction}

There is a striking similarity between two apparently unrelated 
problems: the superconducting-normal phase boundary of a square 
superconducting network (our Fig.~2), and the fractal phase 
boundary (see, for instance, Fig.~6.26 of Ref. \cite{NL}) 
of basins of attraction of a dynamical systems map 
studied last century by Weierstrass and generalized 
much later by Hardy in 1916.

The reason for this very interesting similarity among 
these two apparently unrelated problems is because the 
commensurability condition dominates both problems and 
produces a large dip at $1/2$, and smaller dips at 
$1/4$, $1/3$, $2/5$, etc., as discussed previously 
in this work. 

It is interesting to summarize how to obtain the 
Weierstrass fractal boundary of two basins of 
attraction \cite{NL}.  Consider the dynamical map $M$
$$ (x_{k+1}, \; \theta_{k+1}) = M(x_{k}, \; \theta_{k}) $$ 
defined by, 
$$ x_{k+1} = \lambda x_k + \cos \theta_k \; $$
and
$$ \theta_{k+1} = 2 \theta_k \ ({\rm mod} \; 2\pi) \; .$$

When $1 < \lambda < 2$, the map $M$ has two attractors, at
$x=\pm \infty$.  Indeed, since the eigenvalues of the Jacobian 
matrix are $2$ and $\lambda > 1$, there are no finite attractors. 
Therefore, 
$$ M^N(x_0, \; \theta_0 ) = (x_{N}, \; \theta_{N} \; {\rm mod}(2\pi) ) \; ,$$
and $x_N$ tends to either $+\infty$ or $-\infty$ as $N \rightarrow \infty$, 
except for the unstable boundary set 
$$ x=f(\theta) \; ,$$
for which $x_N$ remains finite.

To locate this $ x=f(\theta) $ boundary set, first note that 
$$ \theta_{k} = 2^k \theta_0 \ ({\rm mod} \; 2\pi) \, .$$
The map is non-invertible since it is two-to-one.
However, any $x_N$ can be selected and then find one orbit 
that ends at $(x_N,\; \theta_N)$, by using the above 
$\theta_k$ and taking 
$$ x_{k-1} = \lambda^{-1} [ x_k - \cos(2^{k-1} \theta_0) ] \; .$$
For a given $(x_N,\, \theta_N)$, this orbit starts at 
$$x_0 = \lambda^{-1} x_N - \sum_{l=0}^{N-1} \lambda^{-l-1} 
\cos(2^l \theta_0) \; .$$
Those $(x_0,\, \theta_0)$ such that $x_N$ is finite as $N \rightarrow \infty$, 
define the boundary $ x=f(\theta) $
between the two basins.  Therefore the relation between 
these $x$ and $\theta$ is given by 

\begin{equation}
 x = - \sum_{l=0}^{\infty} \lambda^{-l-1} \cos(2^l \theta_0) 
\equiv f(\theta) \; .
\end{equation}

This sum obviusly converges, since $\lambda > 1$.  However, its 
derivative 

$$ \frac{df(\theta)}{d\theta} \; = \; \frac{1}{2} \,   
\sum_{l=0}^{\infty} 
\ 
\left( 
\frac{2}{\lambda} \right)^{l+1} \, \sin(2^l \theta) \; .$$

\noindent
diverges, since $\lambda < 2$.  Thus, $f(\theta)$ is non-differentiable.
Like our superconducting-normal phase boundary, it has a large 
cusp at $1/2$, and smaller cusps at $1/3$, $1/4$, $2/5$, etc.  
Moreover, it is also symmetric around $1/2$, and it strongly resembles
the $\Delta T_c(\Phi)$ (obtained near $R(T)=0$) for a square lattice.  
Indeed, $\Delta T_c(\Phi)$ corresponds to $x$, and 
$\Phi$ to $\theta$.

The fractal dimension of $ x=f(\theta) $, Eq.(8) above,  is 

$$d_c = 2 - \frac{\ln \lambda}{\ln 2} \; .$$

The precise value of $d_c$ depends on the value of $\lambda$.
Recall that $1 < \lambda < 2$.
For $\lambda$ sligthly less than two, the fractal dimension 
$d_c$ approaches one, and the dips are not pronounced.  This 
is similar to the superconducting-normal phase boundaries 
measured not too close to $T_c$ (e.g., at mid-point drop for the 
$R(T)$ plot).  When the phase-boundary is measured very near 
$T_c$ (when $R(T)$ is very near zero), the number of discernible 
dips grows and they become very sharp (see, e.g., Figs. 10 and 11 of 
Ref.~\cite{17}).  This would corespond to $\lambda$ slightly above one; thus, 
the fractal dimension $d_c$ of the Weierstrass function 
would be closer to two (i.e., a ``rougher" or ``spikier" curve).

Indeed, Ref.~\cite{17} solved for $\Delta T_c(\Phi)$ beyond the mean field 
theory approximation, obtaining a phase boundary similar to the 
Weierstrass function for $\lambda$ slightly above one, and 
$d_c$ near two.  That superconducting-normal phase boundary 
in ref.~\cite{17} has very sharp cusps and dips, and (like the Weierstrass function) 
it is a {\it phase boundary between attractors}.  The map for 
the superconducting networks is obtained from a real-space 
renormalization-group technique.  The mean field limit provides 
smoother phase boundary with $\lambda$ closer to one.  The 
real-space bond-decimation scheme of Ref.~\cite{17} also favors 
fluxes of the form $\Phi = 2^l \, \Phi_0$.  This is clear from 
the way the real-space renormalization-group scheme is 
constructed, where four elementary cells are ``blocked-away" 
into a larger cell with new renormalized effective couplings.
Four of these super-cells are then blocked away into another, 
larger cell, enclosing $16$ elementary cells (or $4$ supercells).
This process is {\it iterated}, until the renormalization group procedure 
coverges (at the phase boundary) or diverges to fixed points 
located away from the fixed point (e.g., $+\infty$).
This (beyond-mean-field) RG iteration\cite{17} and the Weierstrass 
iteration involve very similar types of maps and this generates 
the strikingly similar curves.

In summary, the Weierstrass function and our real-space renormalization 
group approach\cite{17} both produce phase boundaries 
which are strikingly similar.  In particular, both are non-differetiable, 
symmetric around $1/2$ and have a very similar hierarchy of cusps.


\section{comparison of the phase boundaries of superconducting
honeycomb and kagome networks}

Here we discuss an interesting relation between the phase boundaries of
superconducting honeycomb and kagome networks which is due to the
{\it geometrical\/} arrangements of these two types of lattices. 
Indeed, and as kindly pointed out to us by Y.~Xiao and P.M.~Chaikin, 
it is very useful to focus on the region $0\leq \Phi \leq 1$ for the 
honeycomb network and the region $0\leq \Phi \leq 1/8$ for the kagome network.

As shown in Fig.~8(a) through 8(d), though the overall
shapes of the phase diagrams are different, there is a
one-to-one correspondence between the dips in the honeycomb
$\Delta T_{c}(\Phi)$ for $\Phi$, the flux through an
elementary hexagon, in the range $[0,1]$ and those in the
 kagome $\Delta T_{c}(\Phi)$ for $\Phi$, the flux through an
elementary triangle, in the ranges $[0,\; 1/8]$,
$[1/8,\; 1/4]$, and $[1/4,\; 3/8]$. To state this relationship more precisely,
let $\{p/q\}$ be the set of flux values characterizing a number of
dips in the $\Delta T_{c}(\Phi)$ curve for the honeycomb
network. For instance, as labeled in (a), 
$$\{p/q\} = 1/3,\ 2/5,\ 3/7,\ 1/2,\ 4/7,\ 3/5,\ {\rm and}\ 2/3.$$ 
It is observed that the corresponding set of flux values 
for the dips to occur in the kagome $\Delta T_{c}(\Phi)$ curve
would be $\{p/8q\}$ when $\Phi$ lies in the range $[0, \; 1/8]$.
Similarly, the corresponding sets read, respectively,
$$ \Phi = \{ \frac{1}{8} + \frac{p}{8q} \; = \; \frac{p+q}{8q} \} $$ 
for $\Phi \in [1/8, \; 1/4]$ and
$$ \Phi = \{ \frac{1}{4} + \frac{p}{8q} \; = \; \frac{p+2q}{8q} \} $$ 
for $\Phi \in [1/4,\; 3/8]$.
Note that for $\Phi$ in the range $[1/4, \; 3/8]$, the dips in
the $\Delta T_{c}(\Phi)$ curve become less evident:
only five flux values are observed and labeled.
The location and magnitude of the dips found here 
are consistent with recent very interesting experiments 
by the NEC and Princeton groups \cite{nec1,nec2}.

Recall that kagome magnets are known to have degenerate 
ground states (see, e.g., Ref.~\cite{premi} 
and references therein).
Likewise, for superconducting kagome networks at half filling, 
there are several possible ways to arrange fluxes, 
producing a large degeneracy in the $T=0$ ground state \cite{lin-kagome}.  
%
This issue of degeneracy between two states has been systematically studied 
as a function of temperature via computer simulations on 
superconducting samples with a kagome-arranged 
periodic array of pinning sites \cite{fabiana}. 
The second matching field in this system has two fluxons 
per pinning site.  This corresponds to the $f=1/2$ 
state in the kagome superconducting network. 
For this value of the externally applied magnetic field, 
every hexagon has two states 
(with entropy $k_B \log 2$).
$N$ hexagons would have $2^N$ states, and a very large
entropy 

$$ S^{(N \ {\rm hexagons})} \sim N k_B \log 2 \ .$$  

Thus, at the second matching field, superconductors 
with either a kagome or an hexagonal array of 
pinning sites both have  ``low-energy 
states" with a very large degeneracy and a huge (low-$T$) 
entropy.  Thus, when cooling from high temperatures, it 
is difficult to find a unique $T=0$ ground state. 
Transport measurements and mean field theory perhaps 
might not be sufficient to fully elucidate the 
role of bistability and degeneracy in this system.
In order to explore this scenario in a more 
systematic manner, different tools 
(e.g., flux imaging techniques\cite{tonomura} and computer 
simulations\cite{fabiana} of vortex dynamics 
on kagome lattices) might be needed. 

After this work was completed, we became aware of a very 
interesting relevant work by Park and Huse in Ref.\cite{huse}. 
Using Ginzburg-Landau theory, they study superconducting kagome wire networks 
in a transverse magnetic field when the magnetic flux through an elementary triangle 
is a half of a flux quantum. They calculate the helicity moduli of each phase to
estimate the Kosterlitz-Thouless (KT) transition temperatures. At the KT 
temperatures, they estimate the barriers to move vortices and the effects 
that lift the large degeneracy in the possible flux patterns.

\section{summary}

In conclusion, we present a detailed study of the mean-field
superconducting-normal phase boundaries of superconducting square, honeycomb,
triangular, and kagome networks. Our investigations are based on studying 
the quantum interference effects arising from the summation of all the 
closed paths the electron can take on the underlying lattices.  
Other problems\cite{lin} have also been studied in terms of quantum 
interference of electron paths.
We then adopt a systematic approximation scheme, to obtain the 
spectral edges of the corresponding eigenvalue problems, 
and relate the features in the phase boundaries with the geometry 
of the underlying lattice being explored by the moving electrons.
When the electrons are allowed to explore a sizable region of the 
network, our calculations have quickly reached very close 
convergence to the infinite system size results.
There are two particular advantageous aspects of this approach. First, it
enables us to evaluate the superconducting transition temperature
as a {\em continuous\/} function of the applied magnetic field. Second,
it enables us to achieve a step-by-step derivation of the {\it physical origin\/} 
of the many structures in the phase diagrams---in terms of the 
regions of the lattice explored by the electrons.  In particular, 
the larger the region of the network the electrons can explore 
(and thus more paths are available for the electron), the 
finer structure appears in the phase boundary, and the 
sharper the cusps become.  We find many new interesting features 
in these phase diagrams, which compare well with experiments.

\acknowledgements

We thank Paul M.~Chaikin and Yi Xiao for helpful conversations
and for sending us their experimental results.
Y-LL acknowledges support at West Virginia University
from the National Science Foundation (NSF) under grant No.~DMR-91-20333.
FN acknowledges support from the NSF grant No.~EIA-0130383 and 
from the Frontier Research System, The Institute of Physical 
and Chemical Research (RIKEN), Saitama, Japan.
FN also acknowledges B. Janko for his hospitality during a visit 
at the Materials Science Division of Argonne National Laboratory, 
partially supported by DOE grant No. W-31-109-ENG-38.
Partial support has been also provided by the University of Michigan 
Center for Theoretical Physics, and the Center for the Study of Complex Systems.  

\vfill
\eject

\noindent
* Corresponding author.  E-mail address:  nori@umich.edu

\vspace*{0.1in}
\noindent
** Permanent Address.


\begin{figure}
\caption{The oscillatory phase boundary, $\Delta T_{c}(\Phi)$,
for a single superconducting loop.
The top curve corresponds to a triangle (dashed line), the middle a
square (dotted line), and the bottom a hexagon (solid line).
$\Phi$ is the magnetic flux through these cells in units of $\Phi_{0}$.}
\label{fig1}
\end{figure}

\begin{figure}
\caption{Superconducting transition temperature for the square network
as a continuous function of the applied magnetic field:
$\Delta T^{(n)}_{c}(\Phi)=T_{c}(0)-T_{c}^{(n)}(\Phi)$ versus $\Phi$, the
magnetic flux through an elementary square cell. In (a) we show the
superconducting-normal phase boundaries computed from the truncated
Hamiltonians, $H^{(n)}$, for $\Phi$ in the range between $0.2$ and $0.8$.
We omit the parts of $\Delta T^{(n)}_{c}(\Phi)$ for $\Phi \in [0,\;0.2]$ and
$[0.8,\;1]$ since there are no interesting features in these portions of
$\Delta T^{(n)}_{c}(\Phi)$. From top to bottom, the orders of
truncation are $n=5$ (top curve), $6$, $7$, $8$, $10$, $15$, $23$, $39$,
and $70$. Note the development of fine structures and cusps. The
convergence is monotonic. Note also that the closeness between the curves for
$\Delta T^{(39)}_{c}(\Phi)$ and $\Delta T^{(70)}_{c}(\Phi)$ implies that
$\Delta T^{(70)}_{c}(\Phi)$ has achieved close convergence to the infinite
system size $\Delta T_{c}(\Phi)$. The inset schematically depicts a
square lattice.
In (b), we plot $\Delta T_{c}(\Phi)$ for $\Phi \in [0.2,\; 0.8]$  and
label the values of the magnetic flux where observable cusps/dips occur.
They include $\Phi=1/4$, $2/7$, $3/10$, $1/3$, $3/8$,
$2/5$, $3/7$, $1/2$, $4/7$, $3/5$, $5/8$, $2/3$, $7/10$, $5/7$, and $3/4$.
Here $\Delta T_{c}(\Phi)\equiv \Delta T^{(70)}_{c}(\Phi)
=T_{c}(0)-T_{c}^{(70)}(\Phi)$,
our calculated highest-order approximant.}
\label{fig2}
\end{figure}

\begin{figure}
\caption{Field-dependent transition temperature, $\Delta T_{c}(\Phi)$,
of the superconducting square network for various different ranges of
$\Phi$: from (a) to (g), respectively, $\Phi \in [0,1]$,
$[0.333 \simeq 1/3, \; 0.4765]$, $[0.5235,\; 0.667 \simeq 2/3]$,
$[0.375=3/8, \; 0.3978]$, $[0.4025, \; 0.4286\simeq 3/7]$,
$[0.5714 \simeq 4/7, \; 0.5975]$, and $[0.6022, \; 0.625=5/8]$.
It is clear that (b) is enlarged from the maximum in
the left part of (a) and (c) is enlarged from the maximum in the right part
of (a). Similarly, (d) and (e) are, respectively, the enlargements of the
left and right maxima of (b) while (f) and (g) are, respectively,
the enlargements of the left and right maxima of (c).
We also include the labeling of
the values of $\Phi$ where there are cusps/dips in $\Delta T_{c}(\Phi)$.
For the relations between these sets of flux values in different frames,
see the text. The self-similarity in the phase boundary can be concluded
from the resemblance of these figures though an asymetry in the height
develops in each successive magnification.}
\label{fig3}
\end{figure}

\begin{figure}
\caption{Superconducting transition temperature for the honeycomb network
as a continuous function of the applied magnetic field:
$\Delta T^{(n)}_{c}(\Phi)=T_{c}(0)-T_{c}^{(n)}(\Phi)$ versus $\Phi$, the
magnetic flux through an elementary hexagonal cell. In (a) we show the
superconducting-normal phase boundaries computed from the truncated
Hamiltonians, $H^{(n)}$, for $\Phi$ in the range between $0.3$ and $0.7$.
We omit the parts of $\Delta T^{(n)}_{c}(\Phi)$ for $\Phi \in [0,\;0.3]$ and
$[0.7,\; 1]$ since there are no interesting features in these portions of
$\Delta T^{(n)}_{c}(\Phi)$. From top to bottom, the orders of
truncation are $n=9$ (top curve), $10$, $13$, $16$, $21$, $31$, $41$,
and $104$. Note the development of fine structures and cusps. The
convergence is monotonic. We believe that
$\Delta T^{(104)}_{c}(\Phi)$ has achieved close convergence to the infinite
system size $\Delta T_{c}(\Phi)$. The inset schematically depicts a
honeycomb lattice.
In (b), we plot $\Delta T_{c}(\Phi)$ for $\Phi \in [0.3,\; 0.7]$ and
label the values of the magnetic flux where observable cusps/dips occur.
They include $\Phi=1/3$, $2/5$, $3/7$, $4/9$, $5/11$,
$6/13$, $7/15$, $8/17$, $1/2$, $9/17$, $8/15$, $7/13$, $6/11$, $5/9$,
$4/7$, $3/5$, and $2/3$.
Here $\Delta T_{c}(\Phi)\equiv \Delta T^{(104)}_{c}(\Phi)
=T_{c}(0)-T_{c}^{(104)}(\Phi)$,
our calculated highest-order approximant.}
\label{fig4}
\end{figure}

\begin{figure}
\caption{Superconducting transition temperature for the triangular network
as a continuous function of the applied magnetic field:
$\Delta T^{(n)}_{c}(\Phi)=T_{c}(0)-T_{c}^{(n)}(\Phi)$ versus $\Phi$, the
magnetic flux through an elementary triangular cell. In (a) we show the
superconducting-normal phase boundaries computed from the truncated
Hamiltonians, $H^{(n)}$, for $\Phi$ in the range between $0.15$ and $0.85$.
We omit the parts of $\Delta T^{(n)}_{c}(\Phi)$ for $\Phi \in [0,\; 0.15]$ and
$[0.85,\; 1]$ since there are no interesting features in these portions of
$\Delta T^{(n)}_{c}(\Phi)$. From top to bottom, the orders of
truncation are $n=5$ (top curve), $6$, $7$, $10$, $15$, $29$,
and $60$. Note the development of fine structures and cusps. The
convergence is monotonic and rapid. Note also that the closeness between
the curves for $\Delta T^{(29)}_{c}(\Phi)$ and $\Delta T^{(60)}_{c}(\Phi)$
implies that $\Delta T^{(60)}_{c}(\Phi)$
has achieved close convergence to the infinite
system size $\Delta T_{c}(\Phi)$. The inset schematically depicts a
triangular lattice.
In (b), we plot $\Delta T_{c}(\Phi)$ for $\Phi \in [0.15,\; 0.85]$,
our calculated highest-order approximation to $\Delta T_{c}(\Phi)$, and
label the values of the magnetic flux where observable cusps/dips occur.
They include $\Phi=1/5$, $1/4$, $5/16$, $1/3$, $3/8$, $2/5$,
$5/12$, $3/7$, $7/16$, $4/9$, $9/20$, $1/2$, $11/20$, $5/9$, $9/16$, $4/7$,
$7/12$, $3/5$, $5/8$, $2/3$, $11/16$, $3/4$, and $4/5$.
Here $\Delta T_{c}(\Phi)\equiv \Delta T^{(60)}_{c}(\Phi)
=T_{c}(0)-T_{c}^{(60)}(\Phi)$,
our calculated highest-order approximant.}
\label{fig5}
\end{figure}

\begin{figure}
\caption{Superconducting transition temperature for the kagome network
as a function of the applied magnetic field:
$\Delta T^{(n)}_{c}(\Phi)=T_{c}(0)-T_{c}^{(n)}(\Phi)$ versus $\Phi$, the
magnetic flux through an elementary triangular cell. In (a) we show the
superconducting-normal phase boundaries computed from the truncated
Hamiltonians, $H^{(n)}$, for $\Phi$ in the range between $0$ and $1$.
>From top to bottom, the orders of
truncation are $n=4$ (top curve), $5$, $6$, $8$, $10$, $19$,
and $50$. Note the development of fine structures and cusps. The
convergence is monotonic. Note also that the closeness between
the curves for $\Delta T^{(19)}_{c}(\Phi)$ and $\Delta T^{(50)}_{c}(\Phi)$
implies that $\Delta T^{(50)}_{c}(\Phi)$ has achieved close convergence 
to the infinite system size $\Delta T_{c}(\Phi)$. The inset 
schematically depicts a kagome lattice.
In (b), we plot $\Delta T_{c}(\Phi)$ for $\Phi \in [0,\;1]$ and
label the values of the magnetic flux where observable cusps/dips occur.
They include $\Phi=1/12$, $1/8$, $4/25$, $1/4$, $1/3$, $3/8$,
$5/8$, $2/3$, $3/4$, $19/24$, $7/8$ and $11/12$.
Here $\Delta T_{c}(\Phi)\equiv \Delta T^{(50)}_{c}(\Phi)
=T_{c}(0)-T_{c}^{(50)}(\Phi)$, our calculated highest-order approximant.
Note the absence of the cusp at $\Phi=1/2$. This distinct feature is in sharp
contrast to the cases for square, honeycomb, and triangular networks.}
\label{fig6}
\end{figure}

\begin{figure}
\caption{$\Delta T_{c}(\Phi)$'s as functions of $\Phi$ between $0$ and $1$
for the superconducting square, honeycomb, triangular, and kagome
networks, respectively, from (a) to (d). Notice the difference in the
vertical scales.}
\label{fig7}
\end{figure}

\begin{figure}
\caption{$\Delta T_{c}(\Phi)$ versus $\Phi$. (a) is for the superconducting
honeycomb network for $\Phi$ in the range $[0,1]$. (b), (c), and (d) are for
the superconducting kagome network for $\Phi$, respectively, in
the ranges $[0,\;1/8]$, $[1/8,\;1/4]$, and $[1/4,\;3/8]$.}
\label{fig8}
\end{figure}

\end{document}